\documentclass[twocolumn,
 reprint,
 amsmath,amssymb,
 aps,
 longbibliography,
  prl,
 citeautoscript
]{revtex4}

\usepackage{graphicx}
\usepackage{dcolumn}
\usepackage{bm}
\usepackage{bbold}
\usepackage{amsmath}
\usepackage{verbatim}
\usepackage[applemac]{inputenc}
\usepackage{nameref}
\usepackage{varioref}
\usepackage{hyperref}
\usepackage{cleveref}

\usepackage{color}

\newcommand{\bra}[1]{{\left\langle{#1}\right\vert}}
\newcommand{\ket}[1]{{\left\vert{#1}\right\rangle}}

\newcommand{\avg}[1]{\langle{#1}\rangle}

\begin{document}

\title{Experimental few-copy multi-particle entanglement detection}

\author{Valeria Saggio$^{1*}$}

\author{Aleksandra Dimi\'{c}$^2$}

\author{Chiara Greganti$^{1}$}

\author{Lee A. Rozema$^{1}$}

\author{Philip Walther$^{1}$}

\author{Borivoje Daki\'{c}$^{1,3}$}

\vspace*{-2cm}

\affiliation{\vspace{0.2cm}
$^1$Vienna Center for Quantum Science and Technology (VCQ),
Faculty of Physics, University of Vienna,
Boltzmanngasse 5, Vienna A-1090, Austria\\
$^2$Faculty of Physics, University of Belgrade, Studentski Trg 12-16, 11000 Belgrade, Serbia\\
$^3$Institute for Quantum Optics and Quantum Information (IQOQI), Austrian Academy of Sciences, Boltzmanngasse 3, A-1090 Vienna, Austria
}

\begin{abstract}
Quantum technologies lead to a variety of applications that outperform their classical counterparts. In order to build a quantum device it must be verified that it operates below some error threshold. Recently, because of technological developments which allow for the experimental realization of quantum states with increasing complexity, these tasks must be applied to large multi-qubit states. However, due to the exponentially-increasing system size, tasks like quantum entanglement verification become hard to carry out in such cases. Here we develop a generic framework to translate any entanglement witness into a resource-efficient probabilistic scheme. We show that the confidence to detect entanglement grows exponentially with the number of individual detection events. To benchmark our findings, we experimentally verify the presence of entanglement in a photonic six-qubit  cluster state generated using three single-photon sources operating at telecommunication wavelengths. We find that its presence can be certified with at least $99.74\%$ confidence by detecting $20$ copies of the quantum state. Additionally, we show that genuine six-qubit entanglement is verified with at least $99\%$ confidence by using $112$ copies of the state. Our protocol can be carried out with a remarkably low number of copies, making it a practical and applicable method to verify large-scale quantum devices.
\end{abstract}
\maketitle

\section*{Introduction}
The reliable verification of quantum entanglement \cite{arrazola} is an essential task  for quantum technologies, but it remains a considerable challenge for large-scale quantum systems. The generation of large entangled states \cite{pan10, song, monz, friis, pan18} is required to investigate new quantum phenomena and develop novel applications. At the same time, this makes the problem of reliable verification both more important and significantly more consuming in terms of time and resources. The most exhaustive method for inferring quantum entanglement is to reconstruct density matrices via quantum state tomography \cite{james}. However, the number of measurement settings required to characterize a generic quantum state grows exponentially with the size of the system, making this approach unfeasible for large devices. 
In many cases the full density matrix is not needed and alternative approaches for entanglement detection, such as witness-based methods, have been developed (see \cite{guhnedet} and references therein). 
Although these techniques show significant improvements with respect to the number of measurement settings \cite{tothtwomeas,knips2,tran,bavaresco}, they still require many detection events (i.e. many copies of the quantum state) to extract expectation values of different operators used to construct a witness. Moreover, almost all the standard techniques assume that every detection event is \emph{identical} and \emph{independent}, a situation that is challenging to achieve in practice.
For these reasons, as large quantum devices move closer to practical realization, novel methods are urgently needed that are both reliable and resource-efficient. \\

In the past few years, new approaches exploiting various random sampling techniques have been developed, such as randomized benchmarking \cite{knill07}, quantum state tomography via compressed sensing \cite{gross10} and machine learning \cite{montanaro2,torlai}, direct fidelity estimation \cite{flammia}, self-testing methods \cite{mayers,mckague,banal,millerc,reichardt,mckague2}, quantum state verification \cite{takeuchi,zhu}, entanglement verification \cite{pappa,mccutchen,pallister,schneeloch}, and many others.
Most of these techniques are focused on minimizing the number of measurement settings, while the number of copies of the quantum state needed is still very large. 
In contrast, our goal here is to reliably verify entanglement in a realistic scenario, where the number of copies is finite and rather small. Remarkably, in this case it has been shown in \cite{dimic} that even a single copy of the quantum state can be considered as a meaningful resource for entanglement detection. This improvement is made possible by treating quantum entanglement as the ability of the quantum state to answer certain \textit{yes/no} questions. Given only a finite number of copies, the verification necessarily becomes a probabilistic procedure, that is, a quantum resource is verified only with a certain level of confidence.
In contrast to the standard methods, where the main focus is on the extraction of mean values of operators, the method shown in \cite{dimic} only relies on measurements of a single copy of a quantum state.

Here we extend the method presented in \cite{dimic} to develop a generic framework to translate any entanglement witness into a reliable and resource-efficient procedure and apply it to a real experimental situation. We show that our approach: 
\begin{itemize}
\item[a)] detects entanglement with an exponentially-growing confidence in the number of copies of the quantum state,
\item[b)] is implemented via local measurements only, and
\item[c)] does not require the assumption of \textit{independent and identically distributed} (\emph{i.i.d.}) experimental runs.
\end{itemize}
Furthermore, we show that in certain cases our procedure works even if the number of available copies is less than the total number of measurement settings needed to extract the mean value of the witness operator, i.e. \emph{even if the corresponding witness-based method is not logically possible}. 

We demonstrate the applicability of our method by validating the presence of quantum entanglement in a six-photon cluster state. This state is generated with three high-quality single-photon sources at telecommunication wavelengths and detected with pseudo-number resolving superconducting nanowire detectors. We verify the presence of entanglement with at least $99.74\%$ confidence by using around 20 copies of the quantum state and also show that $112$ copies suffice to certify genuine six-qubit entanglement with at least $99\%$ confidence.
In this way, we lay the foundation for a new efficient and advantageous detection scheme, providing a key tool to characterize quantum devices with minimal resources.

\section*{Probabilistic entanglement verification} 
In the standard witness-based approach, the presence of entanglement is verified by measuring the mean value of the witness operator $W$ to be less than zero, i.e. $\avg W\geq 0$ for any separable state $\rho_{sep}$, where $\avg W = \mathrm{Tr}(W \rho_{sep})$. 
$W$ is in general not locally accessible (one has to decompose it into the sum of local observables $W_k$'s as $W=\sum_{k=1}^LW_k$, where each $W_k$ needs to be measured in a separate experimental run), requiring one to estimate several mean values and therefore demanding a large number of copies. Thus, this technique is not reliable when few copies are available.
Moreover, for a limited number of copies $N$, one has to use $L$ independent measurement settings and ensure that for every individual detection event the source provides exactly the same copy of the quantum state (this is the \emph{i.i.d.} assumption). Neither of these two requirements is very practical.

We overcome both of these difficulties by using a probabilistic framework for entanglement detection.
More precisely, our protocol is centred on a set $\mathcal{M}=\{ M_1, M_2, ... , M_L \}$ of binary local multi-particle observables, which we will show can be derived for any entanglement witness. Each $M_k$ (with $k=1,...,L$) returns a binary outcome $m_k=1,0$, associated with the success or failure of the measurement, respectively.
The procedure consists of randomly drawing the measurements $M_k$'s (each with some probability $\varepsilon_k$) $N$ times from the set $\cal M$ and applying each of them to the quantum state, obtaining the outcomes $m_k$'s.
The set $\cal M$ is tailored such that the probability to obtain success (i.e. to get $m_k=1$ for a randomly chosen $M_k$) for any separable state is upper bounded by a certain value $p_s<1$, that we call \textit{separable bound}. On the other hand, the probability of success is maximized to $p_e$, called \emph{entanglement value}, if a certain entangled state (target state) has been prepared. The entanglement value $p_e$ is strictly greater than the separable bound $p_s$, i.e. the difference $\delta_0=p_e-p_s>0$. 
In a realistic framework, we can prepare a certain state $\rho_{exp}$ and assume that the application of the $M_k$'s on it returns $S$ successful outcomes. The observed deviation from the separable bound $\delta=p_{\mathrm{obs}}-p_s$ (where $p_{\mathrm{obs}}$ is the observed entanglement value) therefore reads
\begin{equation}
\delta =\frac{S}{N}-p_s.
\label{delta3}
\end{equation}
It has been shown in \cite{dimic} that the probability $P(\delta)$ to observe $\delta>0 $ for any separable state is upper bounded as $P(\delta) \leq e^{-D( p_s+ \delta || p_s ) N}$, which goes exponentially fast to zero with the number of copies $N$. Here $D(x|| y)= x \log \frac{x}{y} + (1-x) \log \frac{1-x}{1-y}$ is the Kullback-Leibler divergence.
Therefore, the confidence $C(\delta)$ of detecting quantum entanglement is lower bounded by $C_{\min}(\delta)$ as follows:
\begin{equation}
C(\delta) =1- P(\delta)\geq1-e^{-D( p_s+ \delta || p_s )N}=C_{\mathrm{min}}(\delta),
\label{confidence2}
\end{equation}
and converges exponentially fast to unity in $N$. From \eqref{confidence2} we can estimate the average number of copies $N_{av}$ needed to achieve a certain confidence $C_0$, meaning that for a target state preparation we find
\begin{equation}\label{Nav}
N_{av}\leq-K\log(1-C_0)=N_{\max},
\end{equation}
which grows logarithmically at the rate of $K=D( p_s+\delta_0 || p_s )^{-1}$ as $C_0$ approaches unity.

\begin{figure}[h]
\centering
\includegraphics[width=8.9cm]{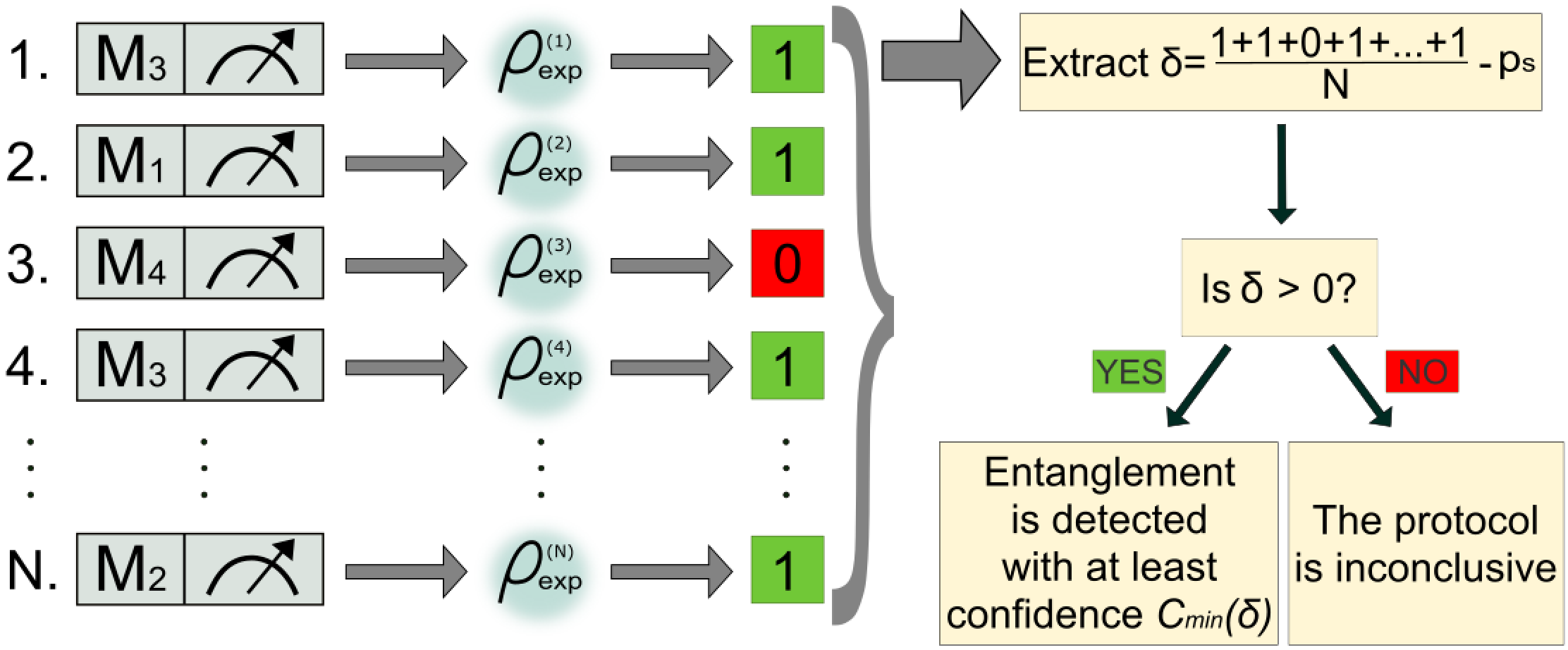}
\caption{\textbf{Protocol for entanglement detection.} The measurements $M_k$'s are randomly sampled from the set $\cal M$ and applied to the experimental state $\rho_{exp}$, which returns binary outcomes $1$ or $0$ (success or failure, respectively). The superscripts in $\rho_{exp}$ account for possible variations of the state due to experimental imperfections. After $N$ runs, the protocol returns $S$ successful outcomes. If the deviation $\delta=S/N-p_s>0$, entanglement is verified in the system with a confidence of at least  $C_{\mathrm{min}}(\delta)$. Otherwise, the protocol is inconclusive.}
\label{meas}
\end{figure}

If $\delta$ 
evaluates to a positive number, we can use \eqref{delta3} to calculate $C_{min}(\delta)$ from \eqref{confidence2}.
We summarize the entanglement detection procedure in Fig.\ \ref{meas}.
\vskip 0.1cm

\vspace{0.4cm}

Additionally, due to random sampling of the measurement settings, our protocol does not require the \emph{i.i.d.} assumption (see \cite{dimic} for the proof). This is an important feature of our procedure as the experimental state is necessarily subject to variations over time due to experimental conditions such as source drift etc. It is known that in such cases other schemes can lead to 
inadequate results \cite{jungnitsch, blume}, whereas in our case we never obtain false positives.

\subsection{Translating entanglement witnesses into the probabilistic framework}
Any entanglement witness 
can be translated into our probabilistic verification protocol. In particular, we will show how to construct the set $\cal M$ and find the corresponding separable bound $p_s$ for any entanglement witness (see Methods, Section I for the detailed proof).
We start with the simple observation that for every witness $W$, one can define a new equivalent one $W' $, whose mean value is always positive and bounded by $1$, by using the equivalence transformation $W' =aW+b$.

The mean value of this new witness 
is the probability of success of our protocol, which is upper bounded by $p_s$ for any separable state and achieves $p_e > p_s$ for a certain entangled state. 
To illustrate the translation procedure, we consider the example of multipartite entanglement detection in an $n$-qubit graph state $\ket{G}$ via the witness $W=\frac{1}{2}\openone-\ket{G}\bra{G}$, for which we have $\avg W\geq0$ for any separable state. This witness can be easily transformed into the equivalent one 
$W'=\frac{1}{2}\openone+\frac{1}{2}\ket{G}\bra{G}$, for which we get $\avg{W'}\leq3/4=p_s$ for any separable state. The graph state can be decomposed as the sum of its stabilizers $S_k$'s as $\ket{G}\bra{G}=\frac{1}{2^n}\sum_{k=1}^{2^n} S_k$, where the $S_k$'s are certain products of local Pauli observables. 
Therefore, the new witness reads $W'=\frac{1}{2^n}\sum_{k=1}^{2^n} M_k$, where $M_k=(\openone+S_k)/2$ are the binary observables needed in our probabilistic protocol. The sampling is uniform, i.e. the probabilities equal $\varepsilon_k=1/2^n$. 
As the $S_k$'s stabilize the state, $p_e=1$ for an ideal graph state. 
We show in the Methods, Section II that this procedure also leads to an estimate of the fidelity $F=\bra{G}\rho_{exp}\ket{G}$ between the experimentally generated state $\rho_{exp}$ and the ideal one $\rho_{ideal}=\ket{G}\bra{G}$. This is related to the direct fidelity estimation protocol \cite{flammia}.

Given $p_e$ 
and $p_s$ we can obtain the average number of copies needed to achieve a certain confidence $C_0$ from \eqref{Nav}. We get $N_{av}\leq -D( 1|| 3/4 )^{-1}\log(1-C_0)\approx -3.48\log(1-C_0)$. Therefore, to achieve confidence of $C_0=0.99$ we need at most $N_{\max}\approx 16$ copies of $\ket{G}$, which is a remarkably low number. Furthermore, this number is independent of the system size (i.e. number of qubits $n$).

Once we have found the $M_k$'s and $p_s$, we can apply the protocol illustrated in Fig.\ \ref{meas} and find the minimum confidence for entanglement detection.\\




The previous example of the graph state shows a constant gap between $p_s$ and $p_e$ that does not depend on $n$. For this reason, the number of required copies needed to achieve a certain confidence does not grow with the number of qubits (we recall that only $16$ copies are required to achieve $99\%$ confidence, regardless of the number of qubits). In this case, the standard witness-based approach would require $2^n$ measurement settings, and each setting would demand a large number of copies, whereas our procedure provides reliable detection with a constant overhead. Thus, our method applies even if the number of settings exceeds the number of available copies. A further reduction of copies (even to a single one) was shown for certain classes of large multi-qubit states \cite{dimic}. More precisely, in \cite{dimic} examples were presented with $p_s=e^{-\alpha n}$ (where $\alpha$ is a constant), which vanishes exponentially fast in $n$, while maintaining $p_e$ constant in $n$. In this case, we can approximate $K\approx1/(\alpha n)$, thus even a single copy of the quantum state suffices to verify entanglement with high confidence (provided that $n$ is sufficiently large). On the other hand, as long as $\delta_0$ does not vanish when increasing the system size, we still have exponential efficiency of the procedure at the constant rate $K$. 
Finally, an interesting case occurs if $\delta_0$ approaches zero as we increase the number of qubits. In this case, we can approximate $K\approx \frac{2p_s(1-p_s)}{\delta_0^2}$, leading to 
$N_{\max}\approx-\frac{2p_s(1-p_s)}{\delta_0^2}\log(1-C_0)$. Therefore, as long as $\delta_0^{-2}$ grows moderately in $n$, the procedure remains resource-efficient as the size of the system grows.

\subsection{Entanglement verification tailored for a six-qubit cluster state}
\label{tailored}
\begin{figure}[h]
\centering
\includegraphics[width=3.6cm]{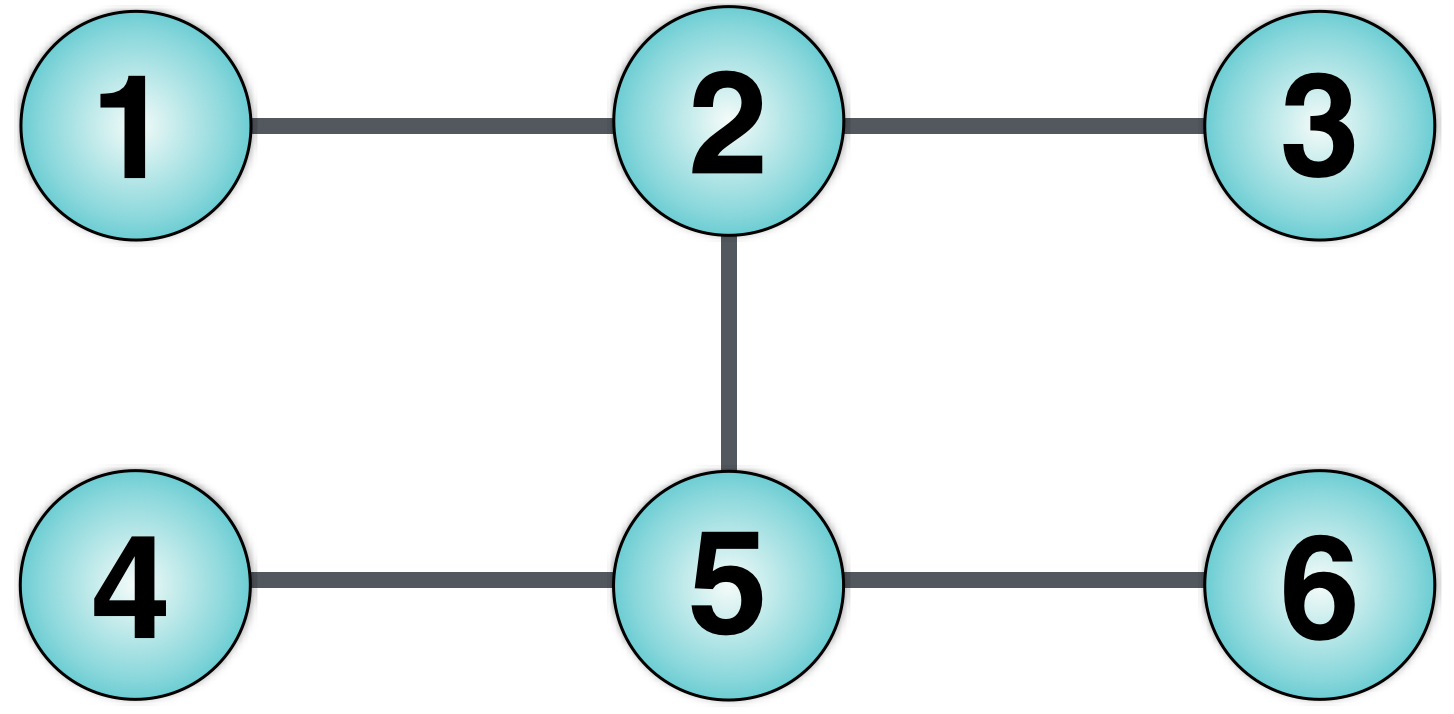}
\caption{\textbf{H-shaped six-qubit cluster state.} Each disk is a qubit prepared in the eigenstate $|+\rangle$ of the Pauli operator $X$, and the solid lines connecting the qubits represent entanglement between them. The entanglement is generated from the application of controlled phase gates between the connected qubits.} 
\label{h1}
\end{figure}

We will now translate two different 
witnesses, tailored for our experimental state, into our probabilistic framework.
Our ideal experimental six-qubit cluster state is
\begin{equation}
\begin{split}|Cl_6 \rangle =& \dfrac{1}{2}(| H_1 H_2 H_3 H_4 H_5 H_6 \rangle + | H_1 H_2 H_3 V_4 V_5 V_6 \rangle + \\ & | V_1 V_2 V_3 H_4 H_5 H_6 \rangle - | V_1 V_2 V_3 V_4 V_5 V_6 \rangle),
\end{split}
\label{cl6}
\end{equation}
which is equivalent to the state shown in Fig.\ \ref{h1} up to local unitary transformations. 

\begin{figure*}[ht]
\centering
\includegraphics[width=6.9in]{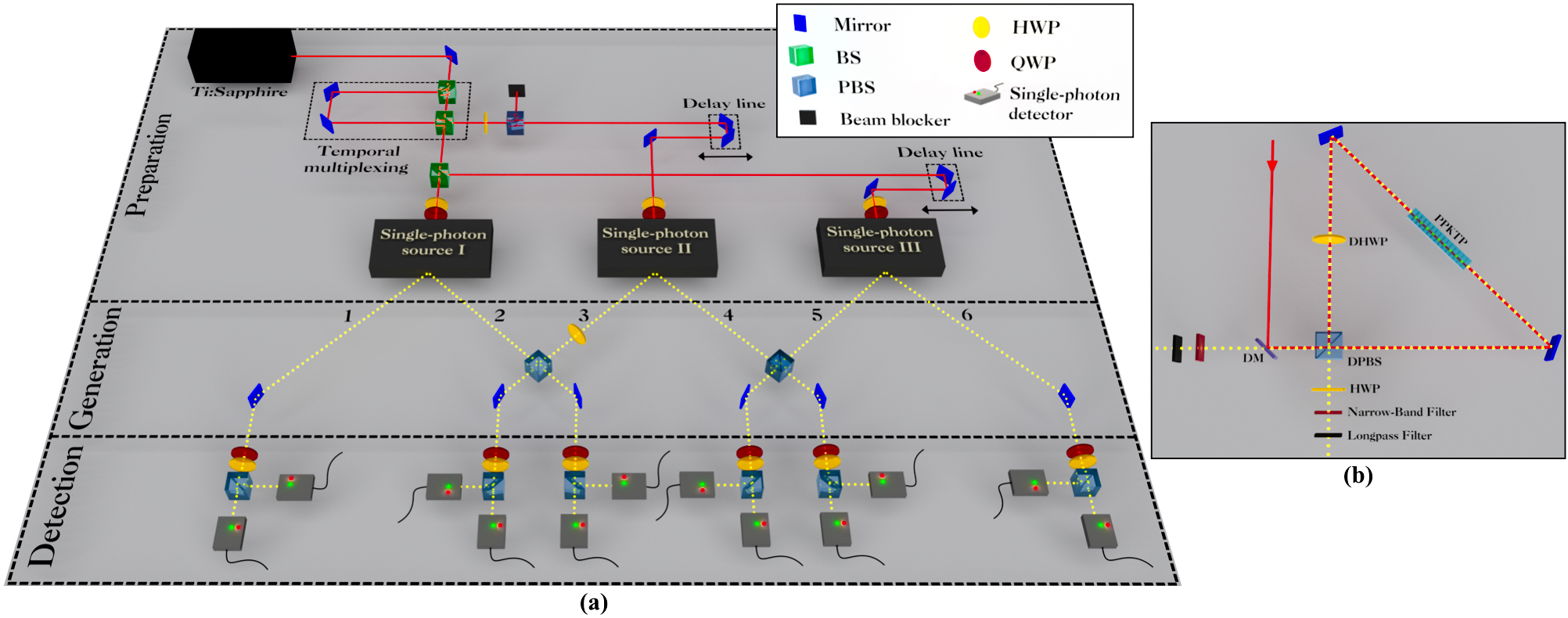}
\caption{\textbf{Experimental setup.} \textbf{(a)} A picosecond Ti:Sapphire laser outputs a beam that is temporally multiplexed to double the repetition rate and reduce contributions from unwanted SPDC high-order emissions. Two beams, equally split at the third BS, pump the first and third single-photon source, while the beam exiting the right output of the second BS passes through a HWP and a PBS before pumping the second source. In this way the power of the second source can be tuned. Movable translation stages are used as delay lines for temporal synchronization. A HWP and a QWP are placed along each beam path to set the needed polarization. Each beam pumps a single-photon source, which emits a polarization-entangled photon pair via type-II SPDC. At each PBS, two photons from different sources interfere. All the photons are then sent to a tomographic system composed of a QWP, a HWP and a PBS. Eventually, photons exiting both outputs of the PBSs reach the single-photon detectors. \textbf{(b)} Schematic of a single-photon source. A PPKTP crystal placed into a Sagnac interferometer is used to generate single photons. DM, Dichroic Mirror; DPBS, Dual wavelength PBS; DHWP, Dual wavelength HWP. Narrow-Band and Longpass filters are respectively used to increase the spectral purity of the photons and cut the residual pump.}
\label{expsetup1}
\end{figure*}

\begin{figure*}[ht]
\centering
\includegraphics[width=7.1in]{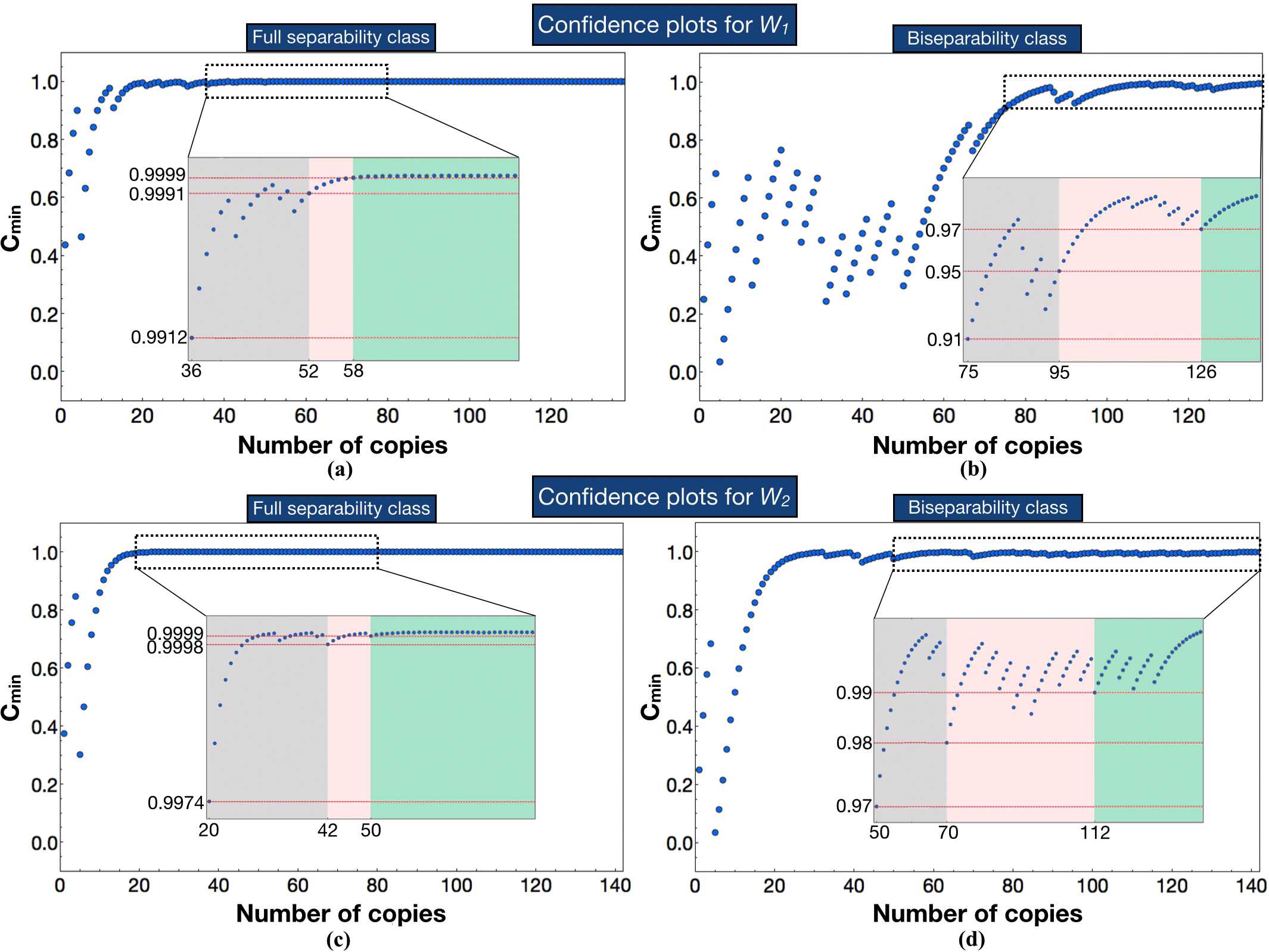}
\caption{\textbf{Growth of confidence of entanglement with the number of copies of the quantum state.} Blue dots represent $C_{\min}$ extracted from (\ref{confidence2}). \textbf{(a)}, \textbf{(b)} show the results for the witness $W_1$, \textbf{(c)}, \textbf{(d)} for the witness $W_2$. \textbf{(a)} and \textbf{(c)} show the minimum confidence when the full separability bound is used (meaning $C_{\min}(S_{W_1}/N-9/16)$ and $C_{\min}(S_{W_2}/N-5/8)$ for \textbf{(a)} and \textbf{(c)}, respectively) and \textbf{(b)}, \textbf{(d)} are extracted by using the biseparable bound (meaning $C_{\min}(S_{W_1}/N-3/4)$ and $C_{\min}(S_{W_2}/N-3/4)$, respectively). $\delta_{W_1}$ and $\delta_{W_2}$ are positive for all the points in the four plots. The region in which the confidence stabilizes is highlighted and shown in the insets, where areas marked with different colors indicate different thresholds for the confidence level. Red dotted lines emphasize the different levels.}
\label{confidencebisep}
\end{figure*}

We consider the two following witnesses, defined to detect genuine six-qubit entanglement:
 \begin{itemize}
 \item[a)] The witness presented in \cite{tothtwomeas}, composed of only two measurement settings: 
 \begin{equation}
 W_1=3\openone-2\left(\prod_{k=1,3,5} \frac{\openone+G_k}{2} + \prod_{k=2,4,6} \frac{\openone+G_k}{2}\right),
 \end{equation}
 \end{itemize}
 where the $G_k$'s (with $k=1,...,6$) are the experimental generators of the cluster state \cite{pan1}, listed in the Methods, Section II; 
 \begin{itemize}
 \item[b)] The standard witness tailored for our cluster state \cite{hein1}: 
 \begin{equation}
 W_2=\frac{1}{2}\openone-\ket{Cl_6}\bra{Cl_6},
 \end{equation}
\end{itemize}
which requires $2^6=64$ measurement settings (since $\ket{Cl_6}\bra{Cl_6}=\frac{1}{2^6} \sum_{k=1}^{2^6} S_k$, analogously to the previous graph state example).

For both witnesses $\avg{W_1}$, $\avg{W_2}\geq0$ for any biseparable state, that by definition does not contain genuine six-qubit entanglement. Nevertheless, both can be also used to distinguish fully separable and entangled states, i.e. to detect some entanglement, and the corresponding separable bounds can be evaluated numerically \cite{gerke}. For this reason, we distinguish two types of separable bounds: one is the so called \textit{biseparable bound} $p_{bs}$, that can be directly extracted from our translation protocol and is therefore used for detection of genuine six-qubit entanglement, the other one is the \textit{full separability bound} $p_{fs}$, which is evaluated numerically and used to detect some entanglement.

Following the procedure shown in the Methods, Section I, we find for $W_1$ the set $\mathcal{M}_{W_1}=\{M_1=\prod_{k=1,3,5} \frac{\openone+G_k}{2}, M_2=\prod_{k=2,4,6} \frac{\openone+G_k}{2}\}$, where $M_1$ and $M_2$ are the binary local observables, and the corresponding biseparable bound is $p_{bsW_1}=3/4$. For $W_2$, the binary observables constituting the set $\mathcal{M}_{W_2}$ are $\frac{\openone + S_k}{2}$ (with $k=1,...,64$) and the biseparable bound is $p_{bsW_2}=3/4$  (see the example of the graph state discussed in the previous section). 
The derived full separability bounds read $p_{fsW_1}=9/16$ and $p_{fsW_2}=5/8$. 
The entanglement values are $p_{eW_1}=p_{eW_2}=1$.

\section*{The experimental setup}\label{expsetup}
The experimental setup used for the cluster state generation is shown in Fig.\ \ref{expsetup1}a. At the \textit{Preparation stage}, a Ti:Sapphire pulsed laser is temporally multiplexed \cite{broome, greganti1} to a repetition rate of $152$ MHz with two \textit{beam splitters} (BSs). 
It then pumps three identical single-photon sources, each built in a Sagnac configuration \cite{kim, fedrizzi, kuzucu, jin}. Each source produces a polarization-entangled photon pair at telecommunication wavelengths via collinear type-II \textit{Spontaneous Parametric Down-Conversion} (SPDC), specifically the singlet state $| \psi ^{-} \rangle_{i,j} = (|H_i V_j\rangle-|V_i H_j\rangle)/\sqrt{2}$, where $|H \rangle,\ |V\rangle$ denote the horizontal and vertical photons' polarization states and $i,j$ the photons' spatial modes. A schematic of one single-photon source is shown in Fig.\ \ref{expsetup1}b (see Methods, Section III for details). It is possible to switch between different Bell states with a \textit{half-waveplate} (HWP) placed along one photon path (see Fig.\ \ref{expsetup1}b) and/or by rotating the HWP positioned along the pump path right before the source.

At the \textit{Generation stage}, after switching from $| \psi^- \rangle_{1,2}$ and $| \psi^- \rangle_{3,4}$ to $| \phi^- \rangle_{1,2}$ and $| \phi^- \rangle_{3,4}$, and from $| \psi^- \rangle_{5,6}$ to $| \phi ^{+} \rangle_{5,6}$, where $| \phi ^{\pm} \rangle_{i,j} = (|H_i H_j\rangle \pm |V_i V_j\rangle)/\sqrt{2}$, photon pairs from different sources interfere at two \textit{polarizing BSs} (PBSs), at which they are temporally synchronized with the help of delay lines placed along the second and third pump paths. A HWP placed in the path of the third photon is needed to generate the target cluster state.

At the \textit{Detection stage}, each photon passes through a tomographic system --- composed of a motorized \textit{quarter-waveplate} (QWP) and HWP followed by a PBS --- that enables measurements in different polarization bases, and is then sent to the detection apparatus, which consists of twelve pseudo-number resolving multi-element superconducting detectors \cite{natarajan, marsili}. Lenses to adjust the beam size, fibers and manual polarization controllers (to compensate for polarization changes into the fibers) are not shown in the figure. 
When the HWP in the third photon path is set to perform a Hadamard gate, 
the simultaneous detection of the six photons at the outputs nominally produces the state \eqref{cl6}.

\section*{Results}\label{results}
For the witness $W_1$, we measured $N_{W_1}=150$ different copies of the state using measurement settings that were randomly sampled from the set $\mathcal{M}_{W_1}$. 
Fig.\ \ref{confidencebisep}a,b show plots of the minimum confidence $C_{\min}(\delta_{W_1})$ versus the number of copies $N$ when the full separability bound $p_{fsW_1}$ and biseparable bound $p_{bsW_1}$  are used, respectively. The points are obtained by plugging the experimentally observed $\delta_{W_1}$ into \eqref{confidence2} to find $C_{\min}(\delta_{W_1})$. 

For the witness $W_2$, we measured $N_{W_2}=160$ different copies of the state using measurements chosen randomly from the set $\mathcal{M}_{W_2}$. As before, Fig.\ \ref{confidencebisep}c,d show the increase in the minimum confidence in the full separability (where $p_{fsW_2}$ is used) and biseparability (where $p_{bsW_2}$ is used) cases, respectively.  
Measurement results from both witnesses that did not register any six-fold coincidence event have not been taken into account (see Methods, Section III).

The experimental plots confirm the efficiency of our entanglement verification method by showing an exponential growth of the confidence. The insets show that the confidence stabilizes towards a certain value with $N$. Since usual technical imperfections lead to experimentally generated multi-qubit states whose fidelities are not perfect, it is expected that the confidence does not show a monotonic growth, because occasional failure events with the binary outcome $0$ will pull the confidence down. Obviously, the fluctuations in the confidence values are linked to the number of measured copies, such that a higher number of copies suppresses these fluctuations. All of this can be seen in Fig.\ \ref{confidencebisep}.

In Fig.\ \ref{confidencebisep}a the confidence stabilizes to at least $99.12\%$ with only $36$ copies. Already $58$ copies suffice to exclude full separability in the system with at least $99.99\%$ confidence. Fig.\ \ref{confidencebisep}b shows verification of genuine six-qubit entanglement with at least $91\%$ confidence with $75$ copies, and already $126$ copies suffice to reach at least $97\%$.

In Fig.\ \ref{confidencebisep}c we see that only $20$ copies suffice to reveal the presence of entanglement with at least $99.74\%$ confidence, and $50$ copies provide more than $99.99\%$. Fig.\ \ref{confidencebisep}d shows that biseparability can be excluded with more than $97\%$ confidence with $50$ copies, and $112$ copies provide more than $99\%$. Interestingly, in this case our protocol works with fewer copies than the total number of measurement settings, i.e. $64$.
As previously discussed, in this last case we can also estimate the fidelity $F=\langle Cl_6 | \rho_{exp} | Cl_6 \rangle = 0.75 \pm 0.06$.
The different areas marked with different colours in both plots and the red dotted lines help the visualization of the different confidence levels.

In our new approach we bypass the measurement of mean values. Our results clearly show that we are able to detect entanglement with a very high confidence using only a few copies of the quantum state.

\section*{Conclusions}
We have provided a generic framework to translate existing witness-based methods for entanglement detection into our probabilistic framework. We have shown that some of the main issues related to standard methods can be overcome when using our protocol. In particular, we have taken a novel approach to the issue of resource efficiency by dramatically reducing the number of copies of the quantum state. 
Combining the novelty of our theoretical model together with advanced technology, including high-purity telecommunication-wavelength multi-photon source and high-efficiency multi-element superconducting detectors, we are able to certify the presence of entanglement in our six-qubit cluster state with at least $99.74\%$ confidence by using only 20 copies of the state and to verify genuine six-qubit entanglement with at least $99\%$ confidence with only $112$ copies in a very short time. 
This approach enables a significant reduction of resources and provides an easy tool to certify the presence of entanglement in large systems. The practicability of our method may prove essential for entanglement detection in large-scale systems in future experiments.



\section{Acknowledgments}
The authors thank I. Alonso Calafell for help with the detectors and T. Str\"{o}mberg for helpful discussions. V.S. acknowledges support from the University of Vienna through the Vienna Doctoral School. A.D. acknowledges support from the project no. ON171035 of Serbian Ministry of Education and Science and from the scholarship awarded from The Austrian Agency for International Cooperation in Education and Research (OeAD-GmbH). L.A.R. acknowledges support from the Templeton World Charity Foundation (fellowship no. TWCF0194). P.W. acknowledges support from the European Commission through ErBeStA (No. 800942), from the Austrian Science Fund (FWF) through CoQuS (W1210-4) and NaMuG (P30067-N36), the U.S. Air Force Office of Scientific Research (FA2386-232 17-1-4011), the Austrian Research Promotion Agency (FFG) through the QuantERA ERA-NET Cofund project HiPhoP, and Red Bull GmbH. B.D. acknowledges support from the Foundational Question Institute (FQXi) grant FQXi-MGA-1806.



\noindent{$^{*}$ Corresponding author, valeria.saggio@univie.ac.at.}


%

\vspace{0.5cm}

\section*{METHODS}

\section*{Section I: Formal proof for generic witness translation}\label{Append1}
Here, we show how to translate any entanglement witness 
into our probabilistic protocol. Conventionally, a witness operator $W$ is normalized such that $\avg{W}=\mathrm{Tr}(W\rho_{sep})\geq 0$ for any separable state $\rho_{sep}$. An equivalent form reads $W=g_{s}\openone-O$, where $O$ is an Hermitian operator for which $\avg O = \mathrm{Tr}{(O \rho_{sep})} \leq g_{s}$ holds for any $\rho_{sep}$ \cite{toth2}. Now, let us consider the local decomposition $O=\sum_{k=1}^q W_k$, where $q$ is the number of local settings needed to measure $\avg O$. We are free to add a constant term to each local component $W_k^{'}= W_k+a\openone$ such that they become non-negative observables. This transformation leads to the new witness $O'=\sum_{k=1}^qW_k^{'}=O+aq\openone$. We choose $a\geq0$ to take the minimum possible value. Altogether, we can rewrite the separability condition as
\begin{equation}\label{sepcond}
\avg{O'} = \mathrm{Tr}(O'\rho_{sep})\leq g_{s}+aq.
\end{equation}
Our main aim is to test this inequality in practice via our probabilistic procedure. Note that this inequality is violated for certain entangled (target) state $\rho_{ent}$, i.e. $\mathrm{Tr}(O'\rho_{ent})=g_e+aq$, with $g_e-g_s>0$. We proceed by writing the spectral decomposition $W^{'}_k=\sum_{s=1}^{\mu_k}\lambda_{ks}M_{ks}$, where $M_{ks}$ are eigen-projectors (binary observables), with $\lambda_{ks}>0$ since $W_k$'s are non-negative operators. The number $\mu_{k}$ counts the non-zero eigenvalues of $W_k$. Furthermore, we define the constant $\tau=\sum_{k=1}^q\sum_{s=1}^{\mu_k}\lambda_{ks}$. We have all we need to set up our verification procedure. As the $W_k$'s are local observables, the binary operators $M_{ks}$'s are local as well. They constitute the set $\mathcal{M}$, which contains in total $L=\sum_{k=1}^q\mu_k$ elements. The probability weights for $M_{ks}$'s are set to $\varepsilon_{ks}=\lambda_{ks}/\tau$.  For a given copy of a separable state $\rho_{sep}$, the probability to obtain success for a randomly drawn measurement $M_{ks}$ from the set $\mathcal{M}$ is given by
\begin{equation}
p=\sum_{k=1}^q\sum_{s=1}^{\mu_k}\varepsilon_{ks}\mathrm{Tr}(M_{ks} \rho_{sep})=\frac{1}{\tau}\sum_{k=1}^q\avg{W_k'}\leq\frac{1}{\tau} ( g_{s}+aq).
\end{equation}
Therefore, the separable bound is given by $p_s=\frac{1}{\tau} ( g_{s}+aq)$. Clearly, for the target state preparation we obtain $p_e=\frac{1}{\tau} ( g_{e}+aq)$ with the strict separation $\delta_0=p_e-p_s=(g_e-g_s)/\tau>0$. Once we have defined the set $\cal M$ and found $p_s$, we can apply the protocol illustrated in Fig.\ \ref{meas} and find the minimum confidence for detecting quantum entanglement.

\section*{Section II: Generators of the six-qubit cluster state and witness decomposition}\label{AppendA}
Our six-qubit cluster state \eqref{cl6} is uniquely defined by the following six generators \cite{pan1}:
\begin{eqnarray}
\begin{split}
G_1 =&\ Z_1 Z_2,~~G_2 =X_1 X_2 X_3Z_5,~~G_3 = Z_2 Z_3\\
G_4 =&\ Z_4 Z_5,~~G_5 =Z_2X_4 X_5 X_6,~~G_6 = Z_5 Z_6,
\label{gi}
\end{split}
\end{eqnarray}
where $X,Y,Z$ are the standard Pauli operators. From this set, we can construct all the products of $G_k$'s, and there are in total $2^6=64$ independent operators which are called stabilizers. 
This witness allows one to combine three of the six generators of the cluster state into one measurement setting, reducing the number of measurement settings from six to two.
To translate the witness $W_1$ (see main text) into our procedure, we start with $O=2\left(\prod_{k=1,3,5} \frac{\openone+G_k}{2} + \prod_{k=2,4,6} \frac{\openone+G_k}{2}\right)$ and $g_s=3$. 
The witness $O$ is already in the spectral form with $M_1=\prod_{k=1,3,5} \frac{\openone+G_k}{2}$ and $M_2=\prod_{k=2,4,6} \frac{\openone+G_k}{2}$ with eigenvalues $+1$, therefore $a=0$. We get $\tau=4$ and the sampling is uniform from the set $\mathcal{M}_{W_1}=\{M_1,M_2\}$. For the biseparable bound we clearly get $p_{bsW1}=3/4$. For full separability, we used the algorithm presented in \cite{gerke} to obtain $p_{fsW1}=9/16$.

The translation procedure for the witness $W_2$ is explained in detail in the main text. For this witness we obtain a biseparable bound of $p_{bsW2}=3/4$. Also in this case, we numerically found the full separability bound to be $p_{fsW2}=5/8$.

\section*{Section III: Experimental details}
We implement the random measurements $M_k$'s with our tomography setup. We only analyze measurement results consisting of six-fold coincidence events. When more than one six-fold event is detected during the same measurement setting, we only use the first coincidence event, to ensure that only one copy of the state is used per measurement.
We will now give a detailed explanation of Fig.\ \ref{expsetup1}a, providing a technical overview of our setup.

\textit{Preparation stage}. A mode-locked Ti:Sapphire Coherent Mira $900$ laser emits pulsed light at a repetition rate of $76$ MHz and at an average power of $1.2$ W. The pulses have a central wavelength of $772.9$ nm and a duration of $2.1$ ps. The first two BSs along the pump path are used to double the repetition rate of the laser and decrease at the same time the power of each pulse, such that unwanted contributions from SPDC higher-order emissions are reduced \cite{greganti1}. This approach is referred to as \textit{passive temporal multiplexing} \cite{broome}. One output of the second BS is sent to a third BS, which equally splits the pump power. The other one passes through a HWP and a PBS, wherein the reflected port is stopped by a beam block. This allows us to adjust the pump power along this path if needed. The two output beams from the third BS and the one from the PBS go through a HWP and a QWP so that polarization can be adjusted, and are then used to pump three single-photon sources. Delay lines in the second and third beam paths are needed later for temporal synchronization. A photon pair is generated from each source via collinear type-II SPDC from a $30$ mm long \textit{periodically poled KTiOPO$_4$} (PPKTP) crystal placed into a Sagnac interferometer, which has the advantages of compactness and phase stability. A schematic of a single-photon source is shown in Fig.\ \ref{expsetup1}b. It is composed of a \textit{dichroic mirror} (DM) reflecting the pump and transmitting the photons, a \textit{dual PBS} (DPBS) and a \textit{ dual HWP} (DHWP), which work for both pump and photon wavelengths, and a PPKTP. The crystal temperature set to $24^\circ$ enables photon wavelength degeneracy at $1545.8$ nm. The photons generated from the crystal pass through ultra-narrow filters with a bandwidth of $3.2$ nm that improve their spectral purity and are eventually coupled into single-mode fibers, not shown in the figure. The residual pump beam is removed using longpass filters. 

\textit{Generation stage}. Each pair of photons coming from different sources is sent to a PBS, at which it has been temporally synchronized using the delay lines discussed above. The photons exit in fibers --- not shown in the figure --- and propagate in free space through the PBSs, before being coupled into fibers again. A HWP placed along the third photon path is used to generate the cluster state. 

\textit{Detection stage}. Photons from each output go to free space again and then pass through a system composed of a motorized QWP and HWP followed by a PBS. They are eventually re-coupled into fibers and sent to a detection system composed of 12 multi-element superconducting detectors. Each multi-element detector is made up of four nanowires on the same chip, allowing for a pseudo-number resolution and a high detection efficiency ($0.87$ on average at around $1550$ nm). The detectors operate at a temperature $0.9$ K. Photon coincidences are registered using a custom $64$-channel time-tagging and logic module.\\

Our six-fold coincidence rate is primarily affected by coupling losses at the \textit{Generation stage} coming from the propagation of the photons in free space through the PBSs before being coupled again into fibers and filter imperfections. As coupling losses are largest in the second source, we doubled the second source pump power by rotating the HWP placed before the PBS at the \textit{Preparation stage} to compensate. Our final six-fold rate is around $0.1$ Hz. To ensure that each measurement detects at least one copy of the state in every basis, we set the measurement time to $40$ seconds. The tomography waveplates are automatized using PCB motors.










\end{document}